\documentclass[aps,amsfonts,superscriptaddress,altaffilletter,nofootinbib,a4paper]{revtex4}
\usepackage{graphicx}
\usepackage{amsmath}
\usepackage[dvips]{color}
\usepackage[latin1]{inputenc}
\usepackage[english]{babel}
\usepackage{epsfig}
\newcommand{\be}{\begin{equation}}
\newcommand{\ee}{\end{equation}}
\newcommand{\bea}{\begin{eqnarray}}
\newcommand{\eea}{\end{eqnarray}}

\begin{document}
\title{Chaplygin DGP cosmologies}
\author{Mariam Bouhmadi-L\'{o}pez}
\email{mariam.bouhmadi@fisica.ist.utl.pt}
\affiliation{Centro Multidisciplinar de Astrof\'{\i}sica - CENTRA, Departamento de F\'{\i}sica, Instituto Superior T\'ecnico, Av. Rovisco Pais 1,
1049-001 Lisboa, Portugal}
\author{Ruth Lazkoz}\email{ruth.lazkoz@ehu.es}
\affiliation{Fisika Teorikoa, Zientzia eta Teknologia Fakultatea, Euskal Herriko Unibertsitatea, 644 Posta Kutxatila, 48080 Bilbao, Spain}
\begin{abstract}
A new class of braneworld models displaying late-time phantom acceleration without resorting to a phantom fluid is presented. In this scenario
expansion is fuelled by dark matter together with some effective dark energy capable  of crossing the phantom divide. 
Unlike a previous proposal of this nature, in these models the effective phantom behaviour remains valid at all redshifts for some choices of the free parameters  of the models. The construction is based on the generalised Chaplygin gas, and the cosmological history interpolates between a standard CDM-like behaviour at early times and a de Sitter-like behaviour  at late times, so no
future singularity is reached.
 \end{abstract}

\maketitle
\section{Introduction}
Few things in Physics these days can be adjectivated as so intriguing
as the reason for cosmological accelerated expansion on large
scales. Observations of distant supernovae inform us of a departure
from linearity in Hubble's law for which the most compelling
explanation seems to be that the Universe is speeding up.

The simplest explanation is the presence of a cosmological constant or an exotic fluid
making more than two thirds of the total matter/energy content of
the Universe \cite{Copeland:2006wr}, but there is also the possibility 
the acceleration is a
consequence of infrared modifications of gravity with respect to
Einstein's theoretical framework like in the induced gravity braneworld scenario \cite{Dvali:2000hr,Deffayet,IGbrane,Sahni:2002dx,Lue:2004za,chilazmarqui,Lazkoz:2006gp,Lazkoz:2007zk,Apostolopoulos:2006si,Bogdanos:2006pf,Zhang:2006at,other1,Maeda:2003ar,Wands,Bouhmadi-Lopez:2004ax}, which contains as a particular case the well-known DGP model \cite{Dvali:2000hr}. At present we are far from being able to discard either interpretation, so in parallel to the
observational developments required to answer the cosmic
acceleration puzzle, considerable work is being done along
the lines of theory also.

As the mathematical side of the problem is concerned, finding out
about the kinematical/dynamical properties of a given  model for
the large scale Universe model starting off from a reconstruction of
H(z) is a well posed and clear-cut problem. In practice, however,
the situation is tougher, as one only has at hand discrete datasets
with disturbances (for various  redshifts) from which one aims at extracting information applicable to all redshift values in a range
of interest. Thus, in order to make progress  it is vital to smooth
out the quantities involved in the description.

Ideally, one would like to 
 do this coarsening using a theoretically grounded
description of the background geometry.
Given that at present there is not a single model for dark energy
which is completely satisfactory, the community keeps showing
interest on new models 
which firstly do not depart substantially from the
characteristics of other dark energy models in the market,  
secondly are likely to give a better fit to the data, 
and are thirdly are less problematic in what theoretical aspects are concerned.

Observational data do not seem incompatible with dark energy with phantom-like behaviour at $z=0$ \cite{Percival:2007yw}. On the other hand, this radical behaviour 
may well be a property acquired by dark energy just recently,  so models which do the so called phantom divide crossing keep gathering attention (for an overview of the problem see Ref.~\cite{Nesseris:2006er}).
Quite a few realisations of this transition have been discussed in the literature. One can first distinguish general relativistic models, and, among, them there are models with an hybrid of
quintessence and phantom dark energy called quintom  \cite{quinfirst,allotherquintom}, models driven by a fluid called hessence models \cite{hessence,allotherhessence}
which 
originated after giving quintom models a turn of the screw, scalar field models with internal degrees of freedom \cite{hu}, k-essence cosmologies \cite{kess}, Chaplygin gas cosmologies \cite{chapcross} and 
models sourced by vector-like dark-energy \cite{Wei:2006tn}. Possibilities beyond general relativity are scalar tensor theories~\cite{sca-ten}, $f(R)$ theories \cite{fR}, Gauss-Bonnet gravity \cite{gauss}, and, finally, explanations which rely somehow on the assumption of the existence of extra dimensions (see for example \cite{Sahni:2002dx,Lue:2004za,chilazmarqui,Lazkoz:2006gp,Lazkoz:2007zk,Apostolopoulos:2006si,Bogdanos:2006pf,Zhang:2006at}).

A set of proposals to achieve that crossing are those made in the context of braneworld models and, in particular, those inspired in the LDGP model \cite{Sahni:2002dx,Lue:2004za}.  Basically, those  authors considered
a 4-dimensional brane universe containing  matter and a cosmological constant on the brane but no phantom fluid \cite{Sahni:2002dx,Lue:2004za}. The phantom-like behaviour was an effect due to the extra 5th dimension. 
A nice feature of the models is the effective screening of the cosmological constant, which results in the possibility that its value can be larger that the one observationally allowed  in the LCDM case.

The LDGP cosmologies \cite{Sahni:2002dx,Lue:2004za} where generalised by replacing the cosmological constant with quiessence; i.e. a fluid with constant 
equation of state. These new models are referred to as QDGP cosmologies \cite{chilazmarqui}.  As a consequence of their more dynamical character, the phantom divide crossing is possible, unlike in the LDGP models\footnote{Other generalisations of the LDGP model aiming at describing a mimicry of the phantom divide crossing in induced gravity braneworld  models can be found in Refs.~\cite{Apostolopoulos:2006si,Bogdanos:2006pf,Zhang:2006at}.}. This rests on the possibility of describing the evolution as driven by an effective dark energy density which together with dust is considered to be the fuel of the geometry. Yet, these models can be criticised on the ground that the effective phantom picture always breaks down at some point as the
effective dark energy density will cease to be positive at a sufficiently high redshifts (no matter the values of the free parameters of the model).

Here we put forward a possible remedy in the form of an alternative generalisation of the LDGP models, i.e. we present a class of models in which the phantom picture
stays valid at all redshifts (given some conditions that can be met easily). This is achieved by taking the dark energy component on the brane to be a Chaplygin gas \cite{Kamenshchik}, so we can call these cosmologies Chaplygin DGP (hereafter CDGP) models. 

Before proceeding, we recall that the Chaplygin gas by itself can behave as a phantom fluid \cite{Bouhmadi-Lopez:2004me} in the standard framework of Einstein gravity. However, this phantom fluid could induce a future singularity \cite{BLGDMM}. We will show that CDGP models also circumvent the future singularity issue that could arise in phantom Chaplygin gas models. Indeed, as we will show, in our case the brane is asymptotically de Sitter.

In order to show how our models do indeed display the features anticipated in this section, we will provide in the following sections  analytical details on our construction, and for the sake of completeness, we will also present numerical examples to illustrate the effect of changing the values of the free parameters of the models.

\section{The model}

We consider a DGP${_-}$ braneworld model\footnote{The DGP cosmology has two different brane solutions, corresponding to the two possible ways of embedding the brane in the bulk \cite{Deffayet,Bouhmadi-Lopez:2004ax}, the self-accelerating solution (or DGP${_+}$ branch) and the DGP${_-}$ branch. The latter is the one we are considering in the present work.\label{DGP+-}} where the dark energy component on the brane is modelled by a generalised Chaplygin gas (GCG); i.e.  $p_{\rm ch}=-A/\rho_{\rm ch}^{\alpha}$, and with an extra CDM component. In induced gravity braneworld models, which contain as a particular case the DGP scenario,   if there is no interaction between the brane and the bulk the energy momentum tensor in conserved on the brane, so the usual conservation for the total matter/energy budget follows:
\begin{equation}
\dot\rho+3H(\rho+p)=0,
\end{equation}
but restricting it further so that conservation holds for the two components separately we have for the CDM bit
 \begin{equation}
\dot\rho_m+3H\rho_m=0,
\end{equation}
which integrates to the usual $\rho_m=\rho_{m0}(1+z)^3$, whereas for the dark energy bit we have  
 \begin{equation}
\dot\rho_{\rm ch}+3H(\rho_{\rm ch}+p_{\rm ch})=0,
\end{equation}
which results in
\begin{equation}
\rho_{\rm ch}=\rho_{\rm ch 0}[A_s+(1-A_s)(1+z)^{3(1+\alpha)}]^{\frac{1}{1+\alpha}}\label{rhocdef}.
\end{equation}
The latter is the expression that supports the interest on Chaplygin cosmologies, as it reflects the fact that the energy density of such fluids interpolates between dust and a cosmological constant  \cite{Kamenshchik}.
As customary the subscript $0$ corresponds to the current value of a given quantity, and, in addition we define $A_s=A/(\rho_{\rm ch 0}^{1+\alpha})$.

Putting it all together, the modified Friedmann equation on the brane reads \cite{Deffayet,Maeda:2003ar,Wands}
\begin{eqnarray}
H/H_0=\sqrt{\Omega_{r_c}+\Omega_m(1+z)^3+\Omega_{\rm ch}\left[A_s+(1-A_s)(1+z)^{3(\alpha+1)}\right]^{\frac{1}{1+\alpha}}}-\sqrt{\Omega_{r_c}},
\label{hub} \end{eqnarray}
where 
\begin{equation}
\Omega_m=\frac{\rho_{m_0}}{3H_0^2}, \quad \Omega_{\rm ch}=\frac{\rho_{\rm ch 0}}{3H_0^2}, \quad \Omega_{r_c}=\frac{1}{4r_c^2H_0^2},
\end{equation}
are the convenient usual dimensionless parameters. 

Throughout the paper we will be assuming $0<A_s<1$ and $1+\alpha>0$, which imply that
 the braneworld goes over at low redshifts to a de Sitter cosmology (see Eq.~(\ref{hub})). 

Similarly to what happened with vacuum energy  in the precursor LDGP model \cite{Sahni:2002dx,Lue:2004za}, we also have a screening effect here; 
the fractional amount of energy in the form of a Chaplygin gas in this CDGP model can be 
larger than in its general relativistic counterpart (which would be obtained on the $\Omega_{r_c}\to 0$ limit).

A consequence of Eq.~(\ref{hub}), which leads to relevant insight on the model, is the constraint
\begin{equation}
\Omega_m+\Omega_{\rm ch}=1+2\sqrt{\Omega_{r_c}}.
\label{constraint}\end{equation}
It implies, on the first place that the region $\Omega_m+\Omega_{\rm ch}<1$ is unphysical (theoretical cuts of this sort are actually quite a source of complication when 
one tries to impose observational constraints \cite{Lazkoz:2006gp,Lazkoz:2007zk}). 
Secondly,  although the brane is spatially  flat, the constraint (\ref{constraint}) can be interpreted in the  sense there is a mimicry of closed FRW universes
in the $(\Omega_m,\Omega_{ch})$ plane.

Now, a relevant question in this context is whether the brane super-inflates, and in order to answer to it the Raychaudhuri equation is of use:
%
%
\begin{equation}\label{dotH}
\frac{\dot{H}}{H_0^2}=-\frac{3(1+z)^3H}{{2(H+H_0\sqrt{\Omega_{r_c}})}}
\bigg\{\Omega_m+\frac{(1-A_s)\Omega_{\rm ch}(1+z)^{3\alpha}}
{\left[A_s+(1-A_s)(1+z)^{3(1+\alpha)}\right]^{\frac{\alpha}{1+\alpha}}}
\bigg\}.
%
\end{equation}
At this point it is in order to stress that the Raychaudhuri equation in braneworld models is modified with respect to standard relativistic models, in such a way that
the modification of both this equation and the Friedmann equation lead to a conservation equation for the matter/energy on the brane which in the absence of interaction between the bulk and the brane coincides with the conservation equation one would have in the relativistic situation. 

Recently, an alternative route to the derivation of 
a cosmological model combining DGP gravity and a Chaplygin gas was taken by \cite{roos}. In this paper, the DGP-like modification of the Friedmann equation was combined with the relativistic form of the Raychaudhuri equation to derive a conservation  equation for the possible fluids confined on the brane, the equation of state of them being the remaining necessary input. This procedure does not correspond to a construction derived from the DGP action (there is though the possibility an action might exist from which
this derivation could be possible, but as this point has not been addressed in \cite{roos}, such model can only be regarded as phenomenological at present).

\begin{figure}[h]
\includegraphics[height=5cm]{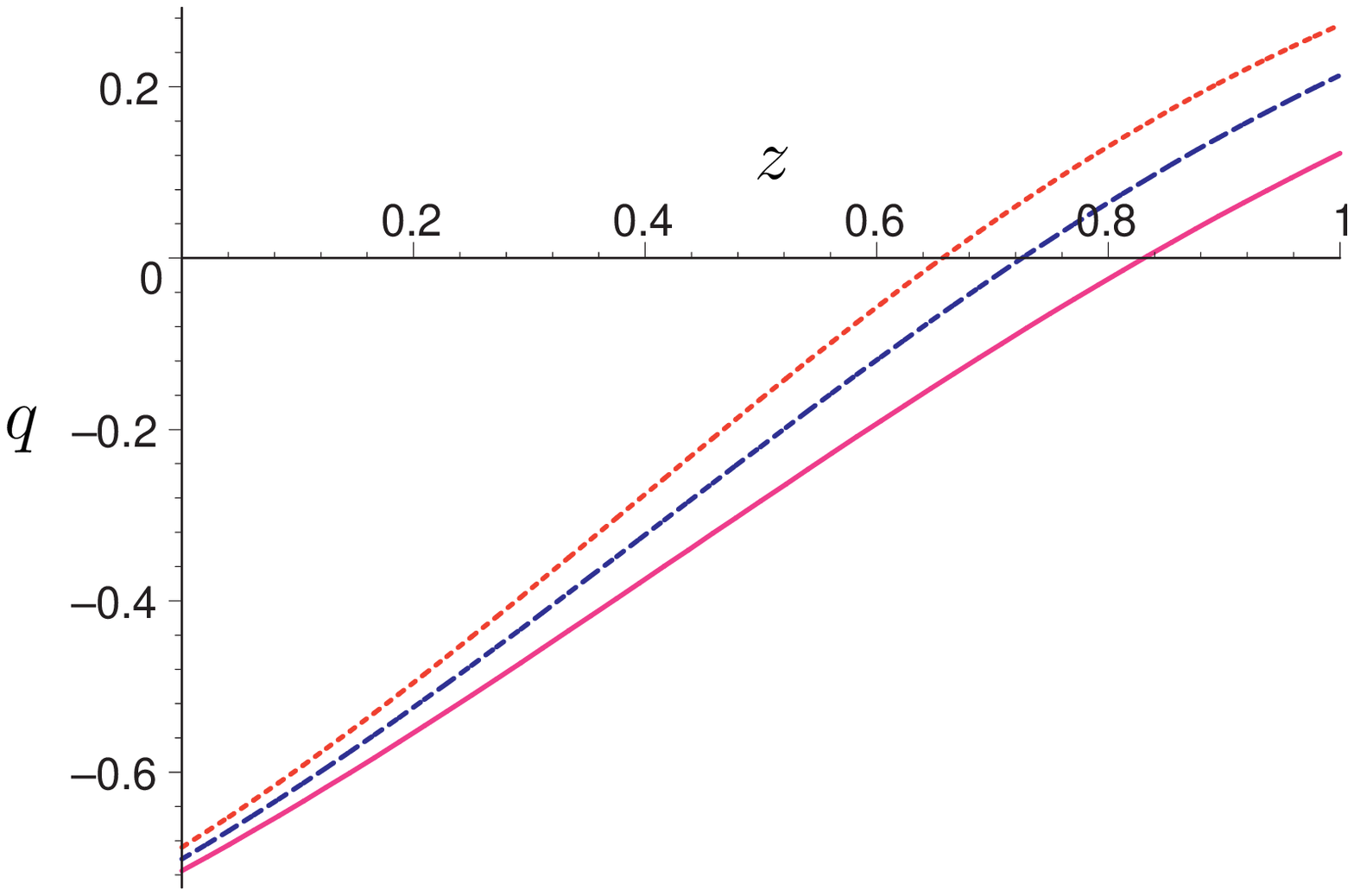} \hspace{0.5cm} \includegraphics[height=5cm]{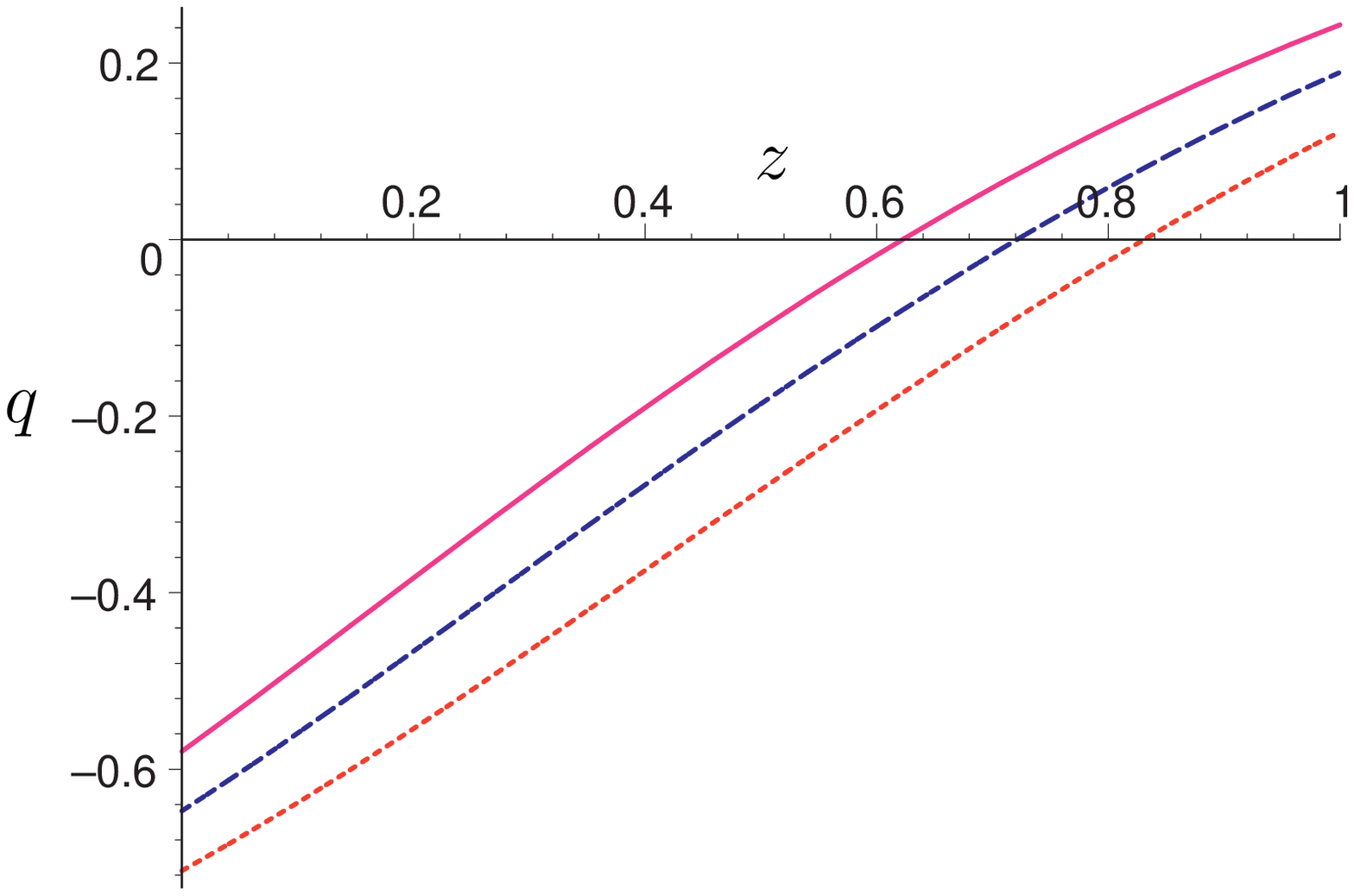}
\caption{Representations of the dimensionless deceleration parameter $q$ against the redshift $z$. The pair of parameters ($\Omega_{r_c}$,$\alpha$) has been fixed to (0.01,0.9). In the left plot $\Omega_m=0.2$,  and the pink solid line, blue dotted line and red dashed-dotted line correspond  to $A_s=0.991,0.981,0.971$, respectively. In the right plot  $A_s=0.991$, and the pink solid line, blue dotted line and red dashed-dotted line correspond  to $\Omega_m=0.3,0.25,0.21$, respectively.}
\label{fig3}
\end{figure}

 In these  models $\dot H$ is never positive (cf. Eq.~(\ref{dotH})),  which means the brane never super-accelerates. Note, as well,  that  $\dot H$   vanishes when $z\rightarrow -1$ while $H$ is positive at this limit, thus reflecting their late-time de Sitter character. In contrast, the large $z$ regime of the models is characterised by $2\dot H=-3H^2$ (this can be deduced by taking the $z\to\infty$ limit in Eqs.~(\ref{hub}) and (\ref{dotH})) . As  the deceleration factor is well known to be 
$q=-(\dot H/H^2+1)$, 
and our brane cosmology transits between  CDM ($q=1/2$) and de Sitter ($q=-1$) limits there is always a certain redshift  below which the brane accelerates (this redshift is signalled by $\dot H = -H^2$). In Fig.~\ref{fig3}, the dimensionless deceleration parameter $q$ is plotted against the redshift $z$ for a fixed set of the parameters $\Omega_{r_c}$ and $\alpha$. More precisely, we show the evolution of $q$ in the left plot of Fig.~\ref{fig3} for a fixed amount of CDM and different values of $A_s$. This plot indicates that the late-time acceleration starts sooner the larger is the parameter $A_s$. On the other hand, we show the evolution of $q$ for a fixed $A_s$ and different amount of CDM in the right plot of Fig.~\ref{fig3}. According to this graph,   the larger is $\Omega_m$, the later the beginning of the cosmic speed up.

By considering that the Universe is currently accelerating, i.e. $\dot H+H^2>0$, and imposing the constraint given in Eq.~(\ref{constraint}) it turns out that 
\begin{equation}
2(\Omega_m+\Omega_{\rm ch})\leq 1+3A_s\Omega_{\rm ch}.
\end{equation}
This implies  the additional constraint
\begin{equation}
\frac{1}{3\Omega_{\rm ch}}\leq A_s
\end{equation}
must be satisfied as well. Notice that the last inequality implies that the Chaplygin gas in this model cannot behave exactly as  dust, i.e. $A_s\neq 0$.

In the next section we will discuss the main point of the paper, which is the possibility of having not just effective phantom behaviour, but also phantom divide crossing. The 
fact this behaviour is effective, i.e. it is achieved without having to resort to scalar fields with a wrong sign in the kinetic energy, allows bypassing the theoretical pathologies these models may display at the quantum level.

\section{Crossing the phantom divide}

As said in the introduction, as well as just above, the phantom behaviour in these CDGP cosmologies is an effective one, and it rests on the definition of  an effective energy density  and 
and effective equation of state parameter $w_{\rm eff}$, so that: 
\begin{eqnarray}
3H^2&=&\rho_m+\rho_{\rm eff},\\
\rho_{\textrm{eff}}&=&\rho_{\rm ch}-3\frac{H}{r_c},\\
1+w_{\textrm{eff}}&=&-\frac{\dot\rho_{\textrm{eff}}}{3H{\rho}_{\textrm{eff}}}.\end{eqnarray}
It can be shown that the effective energy density does not vanish 
and is always positive if
\begin{equation}\label{condition}
4A_s(1-A_s)>\left(\frac{4\Omega_{r_c}\Omega_{m}}{\Omega_{\rm ch}^2}\right)^{1+\alpha}.
\end{equation}
Specifically, the latter condition arises from imposing $\rho_{\rm eff}=0$ and the fulfilment of the Friedmann equation, which translates into a relation between $\rho_m$ and $\rho_{\rm ch}$, which in turn is equivalent to the vanishing of a quadratic polynomial on $(1+z)^{3(1+\alpha)}$. By requiring  such equation has no roots, one just obtains expression (\ref{condition}).
If this  condition is met,  the sign of $1+w_{\textrm{eff}}$ is, therefore, completely determined by the sign of $\dot\rho_{\textrm{eff}}$ or equivalently
by
\begin{equation}
\dot\rho_{\textrm{eff}}=\dot\rho_{\rm ch}-6\sqrt{\Omega_{r_c}}H_0\dot{H}.
\end{equation}
A straightforwards calculation shows
\begin{equation}
\dot\rho_{\rm ch}=-\frac{9HH_0^2\Omega_{\rm ch}(1-A_s)(1+z)^{3(1+\alpha)}}{\left[A_s+(1-A_s)(1+z)^{3(1+\alpha)}\right]^{\frac{\alpha}{1+\alpha}}}.\label{rhoC}
\end{equation}
 By combining Eqs.~(\ref{rhocdef}), (\ref{dotH}) and (\ref{rhoC}), one gets
\begin{equation}
\dot\rho_{\rm eff}=\frac{9HH_0^2(1+z)^3}{(H+H_0\sqrt{\Omega_{r_c}})}\,X
 \end{equation}
where 
\begin{equation}
X=\Omega_mH_0\sqrt{\Omega_{r_c}}-H(1-A_s)\Omega_{\rm ch}(1+z)^{3\alpha}\left[A_s+(1-A_s)(1+z)^{3(1+\alpha)}\right]^{-\frac{\alpha}{1+\alpha}}.
\end{equation}

At high redshifts ($z\gg 1$) the quantity  $\dot\rho_{\textrm{eff}}$ is non-positive as
\begin{equation}
 X\sim -H_0(1-A_s)^{\frac{1}{1+\alpha}} \Omega_{\rm ch}(1+z)^{\frac{3}{2}}\sqrt{\Omega_m+\Omega_{\rm ch}(1-A_s)^{\frac{1}{1+\alpha}}}
.
\end{equation}
Reversely, when $z\approx-1$ the quantity  $\dot\rho_{\textrm{eff}}$ is non-negative,  as one has
\begin{equation}
X\sim \Omega_mH_0\sqrt{\Omega_{r_c}} \quad{\rm for}\;\; \alpha >0.
\end{equation}
Thus, by noticing the opposite signs of $X$ at the two redshift asymptotic regimes  provided $\Omega_m \neq 0$, it follows that this condition is sufficient for  $\dot\rho_{\textrm{eff}}$ to vanish at least once, that is, one is ensuring the realisation of a 
phantom divide crossing. In what follows we will show $\Omega_m \neq 0$ is not only sufficient but also necessary for the passage from the conventional to the phantom regime.

In the absence of CDM  ($\Omega_m=0$) the Raychaudhuri equation reads
\begin{equation}
 2\dot H=-(1+w_{\rm eff})\rho_{\rm eff},
\end{equation}
and since (in all cases) $\dot H\le0$ then $1+w_{\rm eff}\ge0$ and the phantom divide crossing cannot occur. Just for the record, when a Chaplygin gas is advocated to account for both dust and a cosmological constant one speaks about a unified dark matter and dark energy scenario, whereas when the Chaplygin gas is considered in the simultaneous presence of additional dark matter the Chaplygin gas is regarded as a dark energy component. Our findings demonstrate that the Chaplygin gas cannot be the sole component of a DGP braneworld if it is to depict the transition from a non-phantom to a phantom-like regime. This seems to indicate that even though the original inspiration for the use of the Chaplygin gas was the unification of dark matter and dark energy, applications beyond that original motivation are indeed interesting.

Thus, summarising up to here, we can conclude that the phantom divide crossing is realisable in this scenario in the customary effective description, and we want to stress the fact this effective description is well defined at any redshift if  condition (\ref{condition}) holds and $\alpha$ is positive.

It only remains to investigate whether with values of $\Omega_m$ close to observationally preferred ones, and values of $\Omega_{r_c}$ inspired by the closed LCDM ``mimicry'' ($\Omega_{r_c}\ll \Omega_m$) the crossing will occur at observationally accessible values (say $z\le1$ to be a little strict).  This is a tricky task, as  one would has on the one hand the difficulty of having to play with
four parameters  ($\Omega_m. \Omega_{r_c},A_s,\alpha$) and on the other hand the manifest involvement of the pertinent equations. Thus, it does not seem possible to give taxative answers in an analytical way. For that reason, we have resorted to numerical means and, in particular we have done some contour plots to illustrate the effect of changing those parameters. Specifically ,we select constant values of $\Omega_m$ and $\Omega_{r_c}$ and then draw contours of constant $z_c$ (with $z_c$ denoting the redshift value at which the crossing occurs). For the whole range of parameters considered the conditions that the universe is accelerating at present and that $\rho_{\rm eff}$ is positive are satisfied.

\begin{figure}[tbp!]
\includegraphics[height=2.2in]{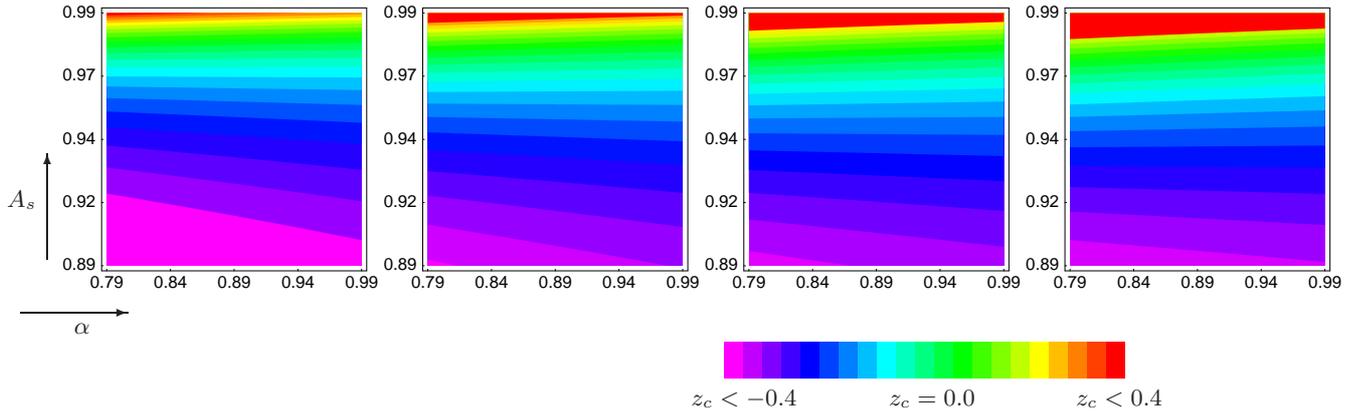} 
\put(-510,50){\vector(0,1){40}}
\put(-520,30){\vector(1,0){40}}
\put(-500,22){$\alpha$}
\put(-525,70){$A_s$}
\put(-269,-5){$z_c<-0.4$}
\put(-195,-5){$z_c=0.0$}
\put(-125,-5){$z_c<0.4$}
\caption{Contour plots of $z_c$ as a function of $\alpha$ and $A_s$ for $\Omega_{r_c}=0.04$ and $\Omega_m=0.20,0.25,0.30,0.35$ (left to right).}
 \label{fig1}\end{figure}

\begin{figure}[tbp!]
\includegraphics[height=2.2in]{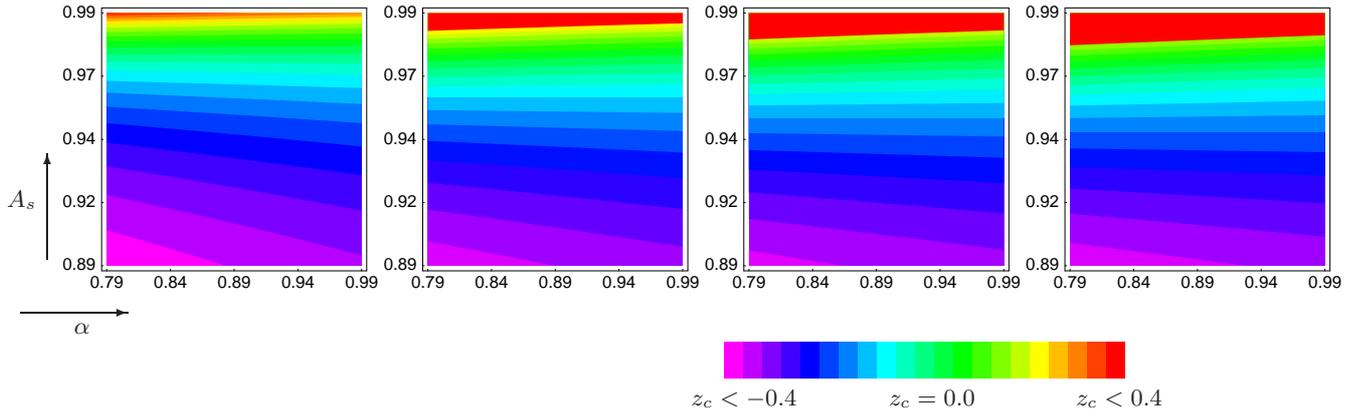} 
\put(-510,50){\vector(0,1){40}}
\put(-520,30){\vector(1,0){40}}
\put(-500,22){$\alpha$}
\put(-525,70){$A_s$}
\put(-269,-5){$z_c<-0.4$}
\put(-195,-5){$z_c=0.0$}
\put(-125,-5){$z_c<0.4$}
\caption{Contour plots of $z_c$ as a function of $\alpha$ and $A_s$ for $\Omega_{r_c}=0.04$ and $\Omega_m=0.25$ and $\Omega_{r_c}=0.01,0.04,0.07,0.10$ (left to right).}
\label{fig2}
 \end{figure}

Our choice of range for $\alpha$ is arbitrary and our only reason for the particular values considered is that the examples  stay close to the original Chaplygin gas ($\alpha=1$). In contrast, our choice of range for $A_s$ is based on some preliminary tests so that our joint choice of values of all four free parameters results in $z_c$ values in the close past. Fig.~\ref{fig1} suggests a pattern according to which for a fixed value of $\Omega_{r_c}$  typically $z_c$ grows with increasing $\Omega_m$, whereas, similarly, in the view of Fig.~\ref{fig2} one would be inclined to say for a fixed value of $\Omega_m$ the  typical $z_c$ values grows with increasing $\Omega_{r_c}$.

\section{Conclusions}

Late-time cosmic acceleration is one of the major issues in Cosmology these days and we have addressed it from the perspective it is an extradimensional effect. We focus on a particular aspect of it, which has aroused a not disdainable activity: the possibility that the equation of state of dark energy (or its effective realisation in our case) has crossed the phantom barrier. 

Here, we present an extra dimensions inspired model that is built on the DGP braneworld scenario \cite{Dvali:2000hr}, which pioneered the attempts of explaining cosmic acceleration initiated at recent times as an infrared modification of gravity \cite{Deffayet,IGbrane}. In broad terms the LDGP \cite{Sahni:2002dx,Lue:2004za} and QDGP \cite{chilazmarqui} variants  of the original DGP$_-$ branch  model (see footnote \ref{DGP+-}) can be viewed effectively as cosmological models filled with CDM and a phantom fluid. The QDGP model presents the advantage with respect to the LDGP model that the phantom divide crossing can be realised (so the effective fluid can be initially conventional as opposed to its latter phantom-like behaviour). However, both models have a drawback: the phantom picture breaks down at sufficiently high redshifts. 

The content of the brane in the LDGP model and the QDGP models is respectively a cosmological constant and quiessence, adding in both cases a CDM component to form the whole matter/energy budget. Here we consider a model with a Chaplygin gas on top of  CDM, and call it CDGP cosmological model. The advantages of this new scenario  are  that (i) the phantom effective picture does not cease to be valid at high redshifts and (ii) a mimicry of the phantom divide crossing is also possible, thus saving some of the problems associated to the LDGP and QDGP models. We also comment briefly on the consistency problems of another Chaplygin DGP model in the literature \cite{roos}.

In the preceding sections, we have presented our model starting off from the Friedmann equation and we have analysed its main evolutionary features. Specifically, we have investigated the conditions for acceleration to occur, and we have concluded that it proceeds provided some restrictions are satisfied by  the free parameters of the model. 
We have also addressed the question of super-acceleration, which we show to be unrealisable as in the LDGP and QDGP models\footnote{See Ref.~\cite{Bouhmadi-Lopez:2005gk} for another variant of the DGP model where super-acceleration is possible and is driven by a bulk scalar field.}.  Another aspect of the problem to which we devote considerable attention is  the role of CDM, as it results crucial for the ability of the model to cross the phantom divide. Finally, we give a numerical account of the redshift values at which the crossing occurs, and with the support of some plots, we show there are values of the parameters for which the passage from the conventional to the phantom-like behaviour can occur at visible redshifts  (see Fig.~\ref{fig1} and \ref{fig2}) while attaining at the same time acceleration at current times (see Fig.~\ref{fig3}).

\section*{Acknowledgements}
M.B.L. is  supported by the Portuguese Agency Funda\c{c}\~{a}o para a Ci\^{e}ncia e
Tecnologia through the fellowship SFRH/BPD/26542/2006 and research grant FEDER-POCI/P/FIS/57547/2004. She also wishes to acknowledge the hospitality of the Theoretical Physics group of the University of the Basque Country during the completion of part of this work.
R.L. is supported
by the University of the Basque Country through
research grant GIU06/37,  and by the
Spanish Ministry of Education and Culture through the
RyC program and research grant FIS2004-01626.

\end{document}